\documentclass[referee]{raa}            
\usepackage{graphicx,times}           \usepackage{natbib}
\usepackage{amssymb,amsmath}
\usepackage[varg]{txfonts}

\usepackage{graphicx}
\usepackage{color}
\usepackage{xspace}
\usepackage{microtype}
\usepackage{mathtools}
\usepackage{multirow}
\usepackage{url}
\usepackage{wasysym}
\usepackage{caption}
\usepackage[version=4]{mhchem}

\def\link_col{blue}
\def\gray{$\gamma$-ray\xspace}
\def\grays{$\gamma$-rays\xspace}
\def\fermi{{\it Fermi}-LAT\xspace}

\def\deg{\hbox{$^\circ$}}

\bibpunct{(}{)}{;}{a}{}{,}

\usepackage[]{hyperref}
\hypersetup{colorlinks=true, linkcolor=green, anchorcolor=red, citecolor=blue, filecolor=red,  urlcolor=red}

\begin{document}

\title{The low cosmic-ray density in Polaris Flare}
\volnopage{Vol.0 (20xx) No.0, 000--000}      
\setcounter{page}{1}          

\author{Zhi-wei Cui
      \inst{1,2}
   \and Rui-zhi Yang 
      \inst{1,2}
   \and Bing Liu
         \inst{1,2,3}
   }
\institute{CAS Key Labrotory for Research in Galaxies and Cosmology, Department of Astronomy, School of Physical Sciences, University of Science and Technology of China, Hefei, Anhui 230026, China\\
\and School of Astronomy and Space Science, University of Science and Technology of China, Hefei, Anhui 230026, China
\and Key Laboratory of Modern Astronomy and Astrophysics (Nanjing University), Ministry of Education, Nanjing 210093, China }

\vs\no{\small Received~~20xx month day; accepted~~20xx~~month day}

\abstract{We reported the \gray observation towards the giant molecular cloud Polaris Flare. Together with the dust column density map, we derived the cosmic ray density and spectrum in this cloud. Compared with the CR  measured locally, the  CR density in Polaris Flare is significantly lower and the spectrum is softer. Such a different CR spectrum reveals either a rather large gradient of CR distribution in the direction perpendicular to the Galactic plane or a suppression of CR inside molecular clouds. }
\authorrunning{}            
\titlerunning{CR in Polaris Flare}  
\maketitle

\section{Introduction}           
\label{sect:intro}
Cosmic rays (CRs) is one of the major composition of  the interstellar medium (ISM). It dominates the heating and ionization of gas inside molecular clouds, regulates the star forming process and  plays a leading role in  astro-chemistry process. It is well believed the CRs are well mixed in the Galactic magnetic field and lost the information of the acceleration site. In this regard, \grays, the secondary products of interactions of CRs with gas and photons in the ISM, provide us crucial information about the distribution and propagation of CRs in the Galaxy.  Thus the giant molecular clouds (GMCs),  regarded as the CR barometers \citep{FA2001}, are the ideal sites to study CRs. The \grays\ in GMCs have already been extensively studied recently, and similar CR density and spectra inside the GMCs in the Gould Belt are derived \citep{neronov12,yang14}.

Polaris flare is a GMC located 240 pc away from the solar system with a total mass of more than $5 \times 10^3 \,\rm M_{\odot}$ \citep{heithausen90}.  The most remarkable feature of this cloud is that there is no any ongoing star-forming process therein. As a result, this cloud can be regarded as an extreme case of the ``passive" clouds in \citet{FA2001}, in which there are no CR injection, and the best place to study the distribution of the Galactic CRs.

In this paper we performed a detailed analysis of \fermi data towards Polaris Flare and derived the CR information therein. This paper is organized as followed: in Sec.2 we derived the gas content in Polaris flare using the Planck dust opacity map, in Sec.3 we describe the \fermi data analysis procedure, in Sec.4 we derived the CR content in this region and discuss the possible implications.

\section{Gas content in Polaris Flare}

\label{sect:gas}

\begin{figure*}[ht]
\centering
\includegraphics[width=0.6\textwidth]{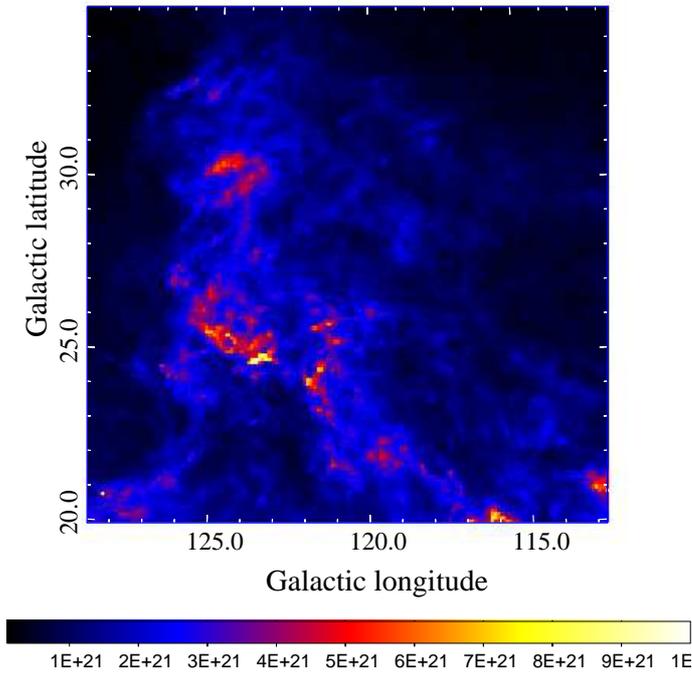}
\caption {The derived total gas column density map from Planck dust opacity for Polaris flare.}
\label{fig:gas}
\end{figure*}

Atomic hydrogen and molecular gas are typically traced by 21-cm \ion{H}{i} line and 2.6-mm CO line, respectively. However, sometimes CO and \ion{H}{i} observations are not enough to account for all the neural gases, especially the ``dark gas"\citep{grenier05}.  Concerning this, we choose the infrared emission from cold interstellar dust as an alternative and independent tracer to estimate the total gas column density.  
According to the research of \cite{planck11}, the dust/gas correlation, i.e. the relation of dust opacity ($\tau_{\rm M}$) as a function of the wavelength $\lambda$ and the total gas column density ($N_{\rm H}$), can be represented by 
\begin{equation}\label{eq:dust}
\tau_{\rm M}(\lambda) = \left(\frac{\tau_{\rm D}(\lambda)}{N_{\rm H}}\right)^{\rm ref}[N_{{\rm H I}}+2X_{\rm CO}W_{\rm CO}].
 \end{equation}
Here $ (\tau_{\rm D}/N_{\rm H})^{\rm ref}$ is the reference dust emissivity measured in low-$N_{\rm H}$ regions, and $X_{\rm CO}=N_{\rm H_{2}}/W_{\rm CO}$ is the  $\rm H_2/CO$ conversion factor, in which $W_{\rm CO}$ is the integrated brightness temperature of the CO emission. Thus, $N_{\rm H}$ as the sum of $N_{\rm HI}$ and $2 N_{\rm H_2}$ is approximated to $N_{{\rm H I}}+2X_{\rm CO}W_{\rm CO}$.
Bring above relations into Equ.\ref{eq:dust}, we get
\begin{equation}\label{eq:dens}
N_{\rm H} =  \tau_{\rm M}(\lambda)\left[\left(\frac{\tau_{\rm D}(\lambda)}{N_{\rm H}}\right)^{\rm ref}\right]^{-1}. 
\end{equation}
Next we set the reference dust emissivity $(\tau_{\rm D}/N_{\rm H})^{\rm ref}=(1.18\pm0.17)\times10^{-26}$~cm$^2$ for the opacity map at $353~\rm GHz$ referring to Table~3 in \cite{planck11}, 
then derive the gas column map for Polaris flare region using Equ.\ref{eq:dens}, which is shown in Fig.\ref{fig:gas}. 

It's worth noting that due to the gradient of the metallicity, the gas-to-dust ratio may also vary in the Galactic scale. For example, the gas-to-dust ratio may be significantly lower \citep{giannetti17} in the Galactic Center than the value derived in \citet{planck11}. 
But it should be acceptable to apply the results in \citet{planck11} for the analysis of nearby sources such as Polaris flare.  

\begin{figure*}[ht]
\centering
\includegraphics[width=0.48\textwidth]{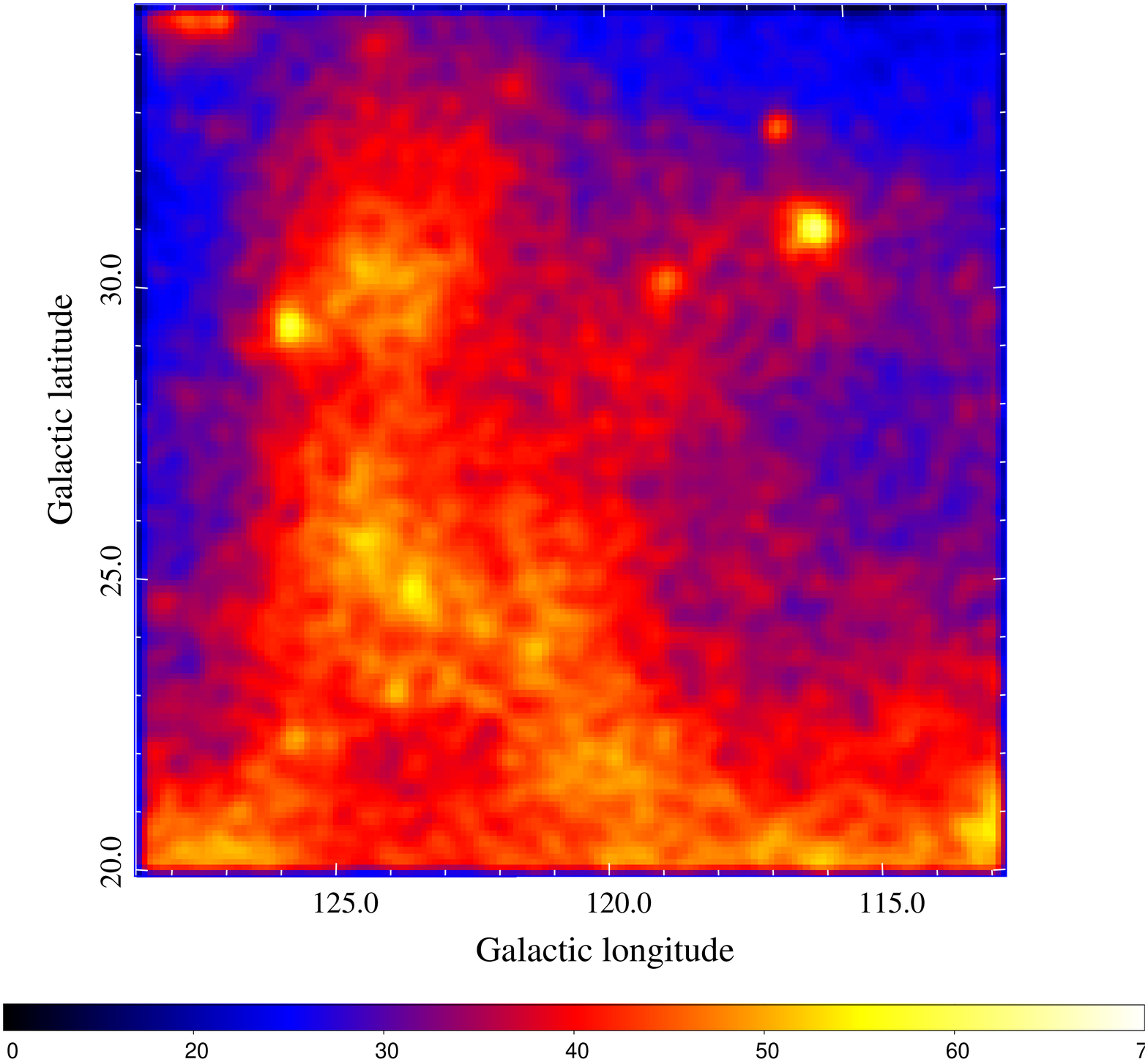}
\includegraphics[width=0.48\textwidth]{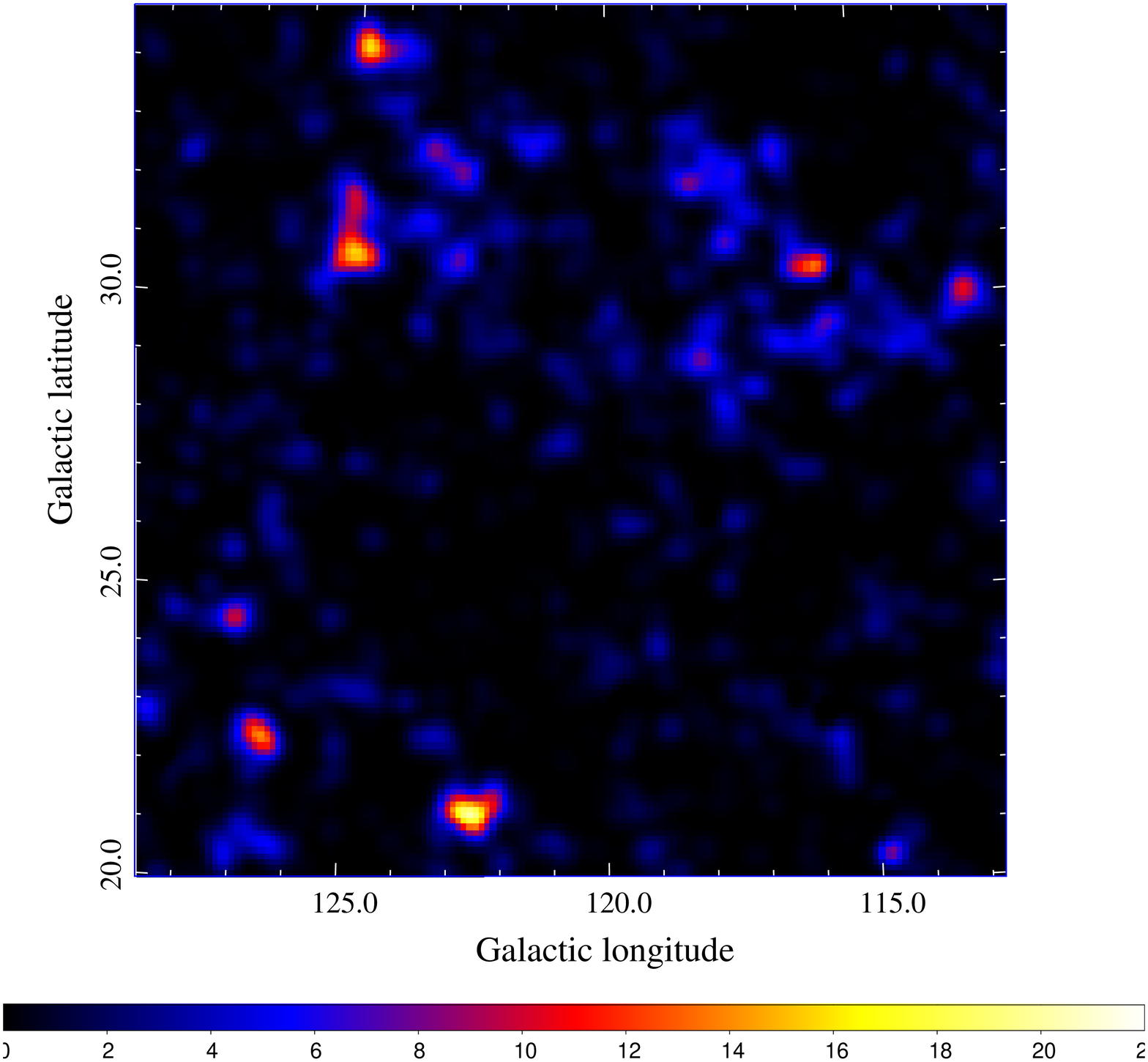}
\caption {The smoothed $15^{\circ}\times15^{\circ}$ counts map of Polaris flare region for \gray  within the energy range of 0.1--300 GeV (left) and the residual TS map  after subtracting all the known sources and diffuse emissions as described in Sec.\ref{sect:data} from the counts map (right).}
\label{fig:cmap}
\end{figure*}

\section{Fermi data analysis}
\label{sect:data}

To study the \gray emission of Polaris flare, we collected 12 years (from 2008-08-04 15:43:36 (UTC) to 2020-08-13 10:30:12 (UTC)) of \fermi Pass 8 data, and used the Fermitools from Conda distribution\footnote{https://github.com/fermi-lat/Fermitools-conda/} together with the latest version of the instrument response functions (IRFs) {\it P8R3\_SOURCE\_V3} for the data analysis.
We chose a { $15^{\circ}\times 15^{\circ}$ }square region centered at the position of Polaris flare (R.A.$=268.021^{\circ}$, Dec.$= 87.961^{\circ}$) as the region of interest (ROI). 
Here we selected the "source" class events, then applied the recommended filter string  ``($\rm DATA\_QUAL>0) \&\& (LAT\_CONFIG==1$)" to choose the good time intervals. Furthermore, to filter out the background \grays from the Earth's limb, only the events with zenith angles less than $90^{\circ}$ are included for the analysis.

The source model for the data analysis, generated by {\sl make4FGLxml.py}, includes the sources in the \fermi 8-year catalog \citep[4FGL,][]{Fermi19} within the ROI enlarged by 7\deg. And for all sources within the ROI, their normalizations and spectral indices are left free. 
However, in order to study the diffuse \gray emissions from the molecular cloud independently,  we did not apply the default \fermi Galactic diffuse background models, i.e. {\it gll\_iem\_v07.fits}, but created our own Galactic diffuse background models. Our diffuse \gray emission model includes two components, one  represents the \grays produced from the pion decay process induced by the inelastic collision between CR protons and ambient gas, and the other represents the \grays resulting from the inverse Compton (IC) scattering of CR electrons in the interstellar radiation fields (ISRFs). The IC component is calculated by GALPROP\footnote{\url{http://galprop.stanford.edu/webrun/}} \citep{galprop}, which uses information regarding CR electrons and ISRFs.

\begin{figure*}[ht]
\centering
\includegraphics[width=0.7\textwidth]{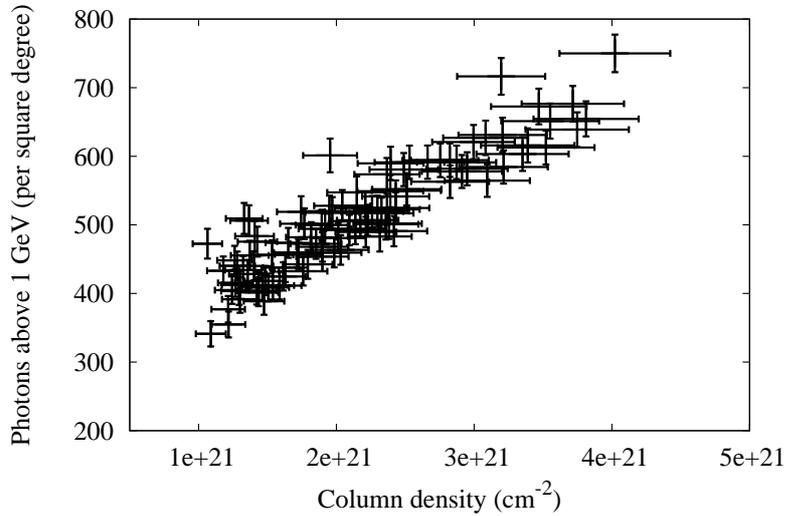}
\caption {Correlation between the photon counts above 1 GeV and the average column density for all pixels with the size of $1\deg\times 1\deg$ within angular distance $<5^{\circ}$ to the center of Polaris flare.}
\label{fig:corr}
\end{figure*}

 We further divided the the pion-decay component into two parts, one associated with the gas the Polaris flare and other associated with the background/foreground gas. We performed such a separation in the gas column density map derived in Sec.2. We attribute all the pixels with column density larger than $2\times 10^{21}~\rm cm^{-2}$ and within $5^{\circ}$ from the center of Polaris flare to the gas associated with the cloud itself, and other pixels to be the background/foreground gas. We note that the exact modeling of the background/foreground gas only have minor effects on the results. This is because in the high latitude region such as Polaris flare, the gas column is dominated by the molecular itself. Indeed, as shown in Fig.\ref{fig:gas}, the average gas column density at the same latitude of Polaris flare ($b \sim 20\deg$) is about $5\times 10^{20} ~\rm cm^{-2}$, which is less than $1/6$ of the average gas column density of Polaris flare itself. To check the correlation of \gray map and gas column density, we plotted the average column density and photon counts above 1 GeV in pixels with the size of $1\deg\times 1\deg$, within the inner $5^{\circ}$ from the center of Polaris flare in Fig.\ref{fig:corr}. We found linear correlation between these two, which reveals that the \grays are indeed associated with the gas in the molecular clouds. Using the modified source model described above, we performed the standard likelihood analysis of \fermi data for events of energies within 0.1--300~GeV. The modified model fits the data very well, as can be seen from the comparison between the counts map of \grays (the left panel of Fig.\ref{fig:cmap}) and the residual TS map after subtract all the known sources and diffuse emissions for Polaris flare region (the right panel of Fig.\ref{fig:cmap}).

\begin{figure}
   \centering
   \includegraphics[width=0.7\textwidth, angle=0]{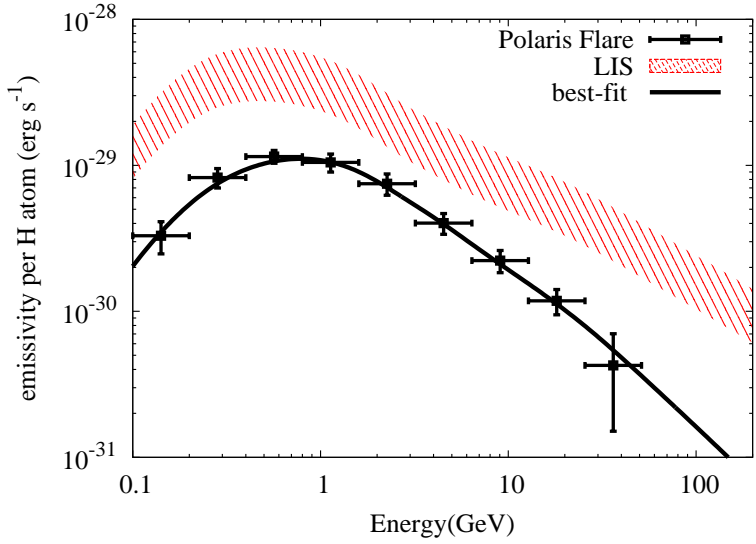}
   \caption{$\gamma$-ray emissivity per H atom derived in Polaris flare. }
   \label{fig:spe}
   \end{figure}

Next, to extract the the spectral energy distribution (SED) of diffuse \grays in this region,  we divided the energy range 100~MeV - 100~GeV into ten logarithmically spaced energy bins, and performed the maximum likelihood analysis on each energy bin.  For each energy bin, the significance of the signal detection exceeds $2\sigma$, and the uncertainties include 68\% statistical errors for the energy flux and the systematic errors due to the uncertainties in LAT effective areas. { As mentioned above the background/foreground gas should cause only minor effects to the results. To investigated the possible systematic errors related to these effects, we firstly varied the selection criterion of the gas associated with the molecular clouds from $1.5\times 10^{21}~\rm cm^{-2}$ to $2.5\times 10^{21}~\rm cm^{-2}$, and then artificially increase or decrease the normalization of the background/foreground gas template in the likelihood fitting by 50\%. We included the uncertainties of the these tests in the error bars in the derived SEDs. } 
Then we further divided the SED with the total gas column densities derived from Sec.\ref{sect:gas} to obtain the \gray emissivities per H atom, which is proportional to the CR density. The results are illustrated in Fig.\ref{fig:spe}. Besides, the derived emissivities are compared with the one predicted by using the local interstellar spectrum (LIS) of CRs (red shaded area, \citealt{casandjian15}).

%
   \begin{figure}
   \centering
   \includegraphics[width=0.7\textwidth, angle=0]{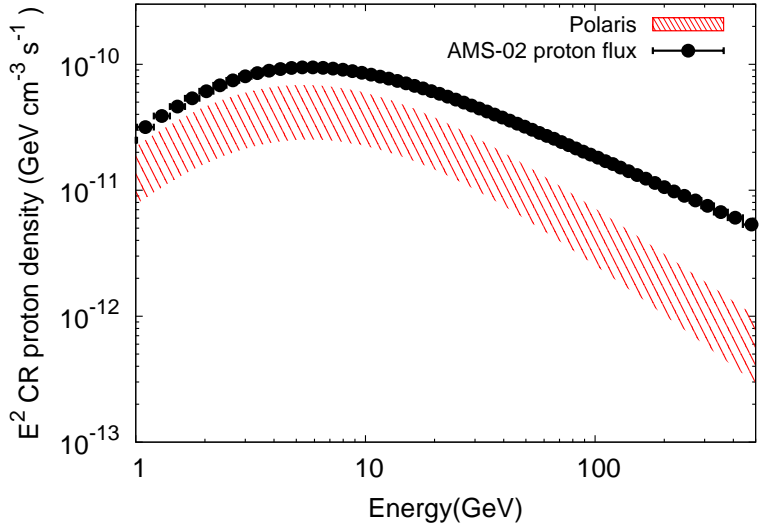}
   \caption{CR density derived in Polaris flare.}
   \label{fig:cr}
   \end{figure}

\section{Discussion and Conclusion}
\label{sect:discussion}
We derived the parent CR proton spectrum by fitting the \gray emissivities using the \gray production cross section in pion-decay process \citep{Kafexhiu14}, in which the nuclear enhancement factor of about 1.8 is also taken into account.  Considering the possible low energy break of CR spectrum, we use the function form $F(E) \sim (E+E_a)^{-\gamma}$.  In deriving the CR density, we have taken into account the uncertainties of about 15\% in the gas-to-dust ratio and about 20\% in the \gray production cross section.  The derived spectral parameters are $\gamma = 3.3 \pm 0.2 $ and $E_a = 3.6 \pm 1.0$, and the corresponding CR spectrum is shown in Fig.\ref{fig:cr}.

We note that the derived CR density is significantly lower (by a factor of 50\% at 100 GeV) than the local measurement. One possible explanation for such low CR density is the lower gas-to-dust ratio in this region, thus the total gas mass are well overestimated and thus the CR density should be higher. The variation of gas-to-dust ratio is predicted due to the variation of metallicity across the Galaxy \citep{giannetti17}. However, due to the proximity of Polaris flare (with a distance of mere 240 pc), the dramatic change of the metallicity and gas-to-dust ratio is very unlikely. 

On the other hand, we note that Polaris flare is about 100 pc above the Galactic plane. The difference in CR density may reflect the gradient of CR distribution perpendicular to the Galactic plane. Indeed, the distribution of CRs along the direction perpendicular to the Galactic plane (z direction) is poorly known. \citet{tibaldo15} has investigated the \gray emission of the high velocity clouds (HVCs) and derived the CR densities in different height above the Galactic plane. But due to the limited size and mass of these HVCs, the uncertainties prevent any strong conclusion.  Theoretically, it depends on the diffusion coefficient ($D_{z}$) along the z direction, as well as the the height ($h$) of the CR halo. In the typical CR propagation models, the escaping of CRs from the Galaxy is dominated by diffusion process. And the confinement time $T \sim \frac{h^2}{D_{z}}$, which can be determined by the secondary to primary ratios. But the absolute value of both $D_{z}$ and $h$ is poorly known so far. If the low CR density in Polaris flare is indeed caused by this effect, the corresponding CR gradient in z-direction would be very dramatic (CR density drops by a factor of two at the height of 100~pc).  Such a strong gradient  is hard to address in the former theoretical calculations for the propagation of CRs in our Galaxy \citep{strong98}.  On the other hand, the large gradient may violate the measured CR anisotropy in the solar neighbourhood. Indeed, the anisotropy can be expressed as $\delta \sim 3D/c\frac{\nabla N}{N}$, where D is the diffusion coefficient, c is the speed of light and N is the CR density.  The measured CR anisotropy at TeV energy is less than $10^{-3}$, thus if the low CR density in Polaris reflects the CR gradient, the upper limit can be derived for the $D_z$. From the equation above, at TeV energy,  $D_z < 0.001 c/3\frac{N}{\nabla N} \sim 0.001 c/3 \frac{100~\rm pc}{0.5}  \sim 6\times 10^{27} ~\rm cm^2/s$, which is significantly lower than expected. 

If this is true, such a decrease in CR density should be also observed in other GMCs in the Gould Belt, which all have a similar height of about 100~pc. But the results in \citet{nero12} and \citet{yang14} do reveal that the CR densities and spectra in other clouds in Gould Belt are similar to the local measurement. Thus the low CR density and soft spectrum in Polaris Flare may be connected with the fact that there is no any star forming activities, thus no any CR injection in this cloud. In this scenario, the CR spectrum measured in the solar system should also be dominated by some nearby sources.  And these nearby sources should have little impact on the CR spectrum in Polaris flare. Because of the proximity of Polaris flare, such a scenario implies that the 100 GeV CRs are not well mixed at the scale of 200 pc. And in the standard diffusion model of Galactic CRs, the diffuse length of CRs can be expressed as $l \sim \sqrt{2DT} \sim 100 {\rm pc} \sqrt{\frac{D}{10^{29} \rm cm^{2}s^{-1}}\frac{T}{10^4 \rm yrs}} $. Thus CRs should be well mixed in this scale unless there are some very recent CR injection events, but such a recent nearby injection event of CRs may already violate the anisotropy measurement.  Another solution would be the much smaller diffusion coefficient than  estimated previously \citep{hawc2017sci}.  Interestingly, such a low diffusion coefficient is also investigated in \citet{qiao21} to fit the CR anisotropy in a broad energy range. 
    
To conclude, we found that the CR density in Polaris flare is significantly smaller than that measured locally, and the spectral shape is also much softer. We argue that there are three possibilities to address such an observational result. One is the variation in the gas-to-dust ratio in Polaris flare, another is the strong CR gradient along the z-direction of our Galaxy, and the third is the slower diffusion of CRs than expected. The latter two cases all require a significant modification to the current understanding of the Galactic CRs. Careful theoretical calculations on the CR propagation and detailed observations on other clouds may help to solve these puzzles.

\begin{acknowledgements}
Rui-zhi Yang is supported by the NSFC under grants 11421303 and the national youth thousand talents program in China. Bing Liu is supported by the Fundamental Research Funds for the Central Universities. 
\end{acknowledgements}

\bibliographystyle{raa}

\bibliography{ms}

\end{document}